\documentclass[aps,prb,reprint,english,american,showpacs,10pt]{revtex4-1}
\usepackage[T1]{fontenc}
\usepackage[latin9]{inputenc}
\usepackage{color}
\usepackage{amsmath}
\usepackage{amssymb}
\usepackage{graphicx}
\usepackage{ulem}

\makeatletter

\newcommand{\noun}[1]{\textsc{#1}}

\@ifundefined{textcolor}{}
{%
 \definecolor{BLACK}{gray}{0}
 \definecolor{WHITE}{gray}{1}
 \definecolor{RED}{rgb}{1,0,0}
 \definecolor{GREEN}{rgb}{0,1,0}
 \definecolor{BLUE}{rgb}{0,0,1}
 \definecolor{CYAN}{cmyk}{1,0,0,0}
 \definecolor{MAGENTA}{cmyk}{0,1,0,0}
 \definecolor{YELLOW}{cmyk}{0,0,1,0}
}

\usepackage[T1]{fontenc}
\usepackage{lmodern}

\renewcommand\[{\begin{equation}}
\renewcommand\]{\end{equation}} 
\renewcommand{\vec}[1]{\mathbf{#1}}

\newcommand{\ve}{\vec e}

\newcommand{\vq}{\vec q}

\renewcommand{\vr}{\vec r}

\newcommand{\vQ}{\vec Q}

\makeatother

\usepackage{babel}
\begin{document}

\title{\textcolor{black}{$\mathbb{Z}_{2}$ phase diagram of three-dimensional
disordered topological insulator via scattering matrix approach }}

\author{Bj\"orn Sbierski}
\author{Piet W. Brouwer}

\affiliation{Dahlem Center for Complex Quantum Systems and Institut f\"ur Theoretische
Physik, Freie Universit\"at Berlin, D-14195, Berlin, Germany}

\begin{abstract}
The role of disorder in the field of three-dimensional
time reversal invariant topological insulators has become an active
field of research recently. However, the computation of $\mathbb{Z}_{2}$
invariants for large, disordered systems still poses a considerable
challenge. In this paper we apply and extend a recently proposed method
based on the scattering matrix approach, which allows the study of
large systems at reasonable computational effort with few-channel leads. By computing the
$\mathbb{Z}_{2}$ invariant directly for the disordered topological
Anderson insulator, we unambiguously identify the topological nature
of this phase without resorting to its connection with the clean case. 
We are able to efficiently compute the $\mathbb{Z}_{2}$ phase
diagram in the mass-disorder plane. The topological phase boundaries
are found to be well described by the self consistent Born approximation, both for vanishing and finite chemical potential.
\end{abstract}


\maketitle

\section{Introduction}

Time reversal invariant (TRI) topological insulators, a class
of insulating materials with strong spin orbit coupling, have 
attracted a great
amount of attention in recent years. While clean systems are fairly
well understood,\cite{Hasan2010,Bernevig2013} an important theme
in current topological insulator research is the study of disorder.
Besides being crucial for the interpretation of experimental data,
disorder is of fundamental interest: Generically, disorder localizes
electron wavefunctions and thus is expected to counteract non-trivial
topology, which, as a global property, requires the existence of extended wavefunctions in the valence and conduction bands.
One of the defining properties of strong topological insulator (STI)
phases is their unusual stability: extended bulk- and gapless edge
electronic states persist for weak to moderately strong disorder.
With increasing disorder strength, the bulk gap gets filled with localized
electronic states, the mobility gap decreases and finally, at the
topological phase transition, the mobility gap closes and the surface
states at opposite surfaces gap out via an extended bulk 
wavefunction.\cite{Shindou2009}

However, disorder physics in topological insulators
is much richer than suggested by the simple scheme above. A drastic
example is provided by the topological Anderson insulator transition,
where increasing disorder drives an ordinary insulator (OI) into a
topologically nontrivial phase.\cite{Li2009,Jiang2009,Groth2009,Guo2010,Guo2010a} Moreover,
the role of different disorder types \cite{Song2012} or spatially
correlated disorder \cite{Girschik2012a} has been addressed in literature.
Further, weak topological insulator (WTI) phases known to be protected
by translational symmetry were shown to be surprisingly stable against
almost all disorder types allowed by discrete 
symmetries. \cite{Ringel2012,Mong2012,Obuse2013}

One of the challenges in the field of disordered topological
insulators is the computation of the $\mathbb{Z}_{2}$ invariants
that characterize strong and weak topological insulator phases.
(Without disorder, the $\mathbb{Z}_2$ invariants can be computed
directly from the band structure.\cite{Fu2007a,Hasan2010,Bernevig2013})
While methods based on exact diagonalization are applicable for two
dimensional systems, their performance for three-dimensional systems
is rather poor\cite{Guo2010,Hastings2011,Leung2012}. For example, a recent study \cite{Leung2012} was
only able to map the $\mathbb{Z}_{2}$ invariant for a few lines in
the disorder strength--Fermi energy plane for a system of 8x8x8 lattice
sites, leaving uncertainties about the possibility to infer 
qualitative and quantitative behavior
in the experimentally relevant thermodynamic limit. 
As an example of an indirect method for calculating the 
$\mathbb{Z}_{2}$ invariant,
the three-dimensional topological Anderson insulator
was argued to be topological nontrivial by employing
the Witten effect.\cite{Guo2010a}
The transfer-matrix method can be used to obtain Lyapunov exponents in a finite-size scaling analysis,\cite{Ryu2012,Yamakage2013a}
which is then used to infer information on topological phase boundaries. 
Drawbacks of this method include 
difficulties in the determination of the phase boundary
between two insulating phases, since size dependence of the decay length
is intrinsically small on both sides of the transition. In the case
of a transition between an insulating topologically trivial and nontrivial
phase, application of open boundary conditions allows for a facilitated
detection of the resulting insulator-(surface)metal transition. However,
this causes a much stronger finite-size effect and renders the interpretation
of the results for finite system sizes rather difficult. 
For example, a recent transfer-matrix
study \cite{Kobayashi2013} speculates about a novel ``defeated WTI''
region in the phase diagram, whose precise nature and properties have not been finally resolved.

As a numerically inexpensive alternative, Fulga \textit{et al.}
proposed to obtain the topological invariants from a topological 
classification of the scattering matrix of a topological 
insulator.\cite{Fulga2012c}
As a Fermi surface quantity, the computational requirements for the
calculation of the scattering matrix scale favorably, so that it is 
accessible with modest effort. The method requires the application of
periodic boundary conditions and considers the dependence of the scattering 
matrix on the corresponding Aharonov-Bohm fluxes. In two dimensions, 
there is only one flux, and the method effectively classifies a 
``topological quantum pump'',\cite{Fu2006,Meidan2010,Meidan2011}
via a mapping similar to that devised by Laughlin to classify the
integer quantized Hall effect.\cite{Laughlin1981}

In
this article, we report on the application of a scattering 
matrix-based
approach to a disordered three-dimensional topological insulator 
model \cite{Zhang2009,Liu2010,Imura2012} that features both strong and weak
topological insulator phases. In Sec. \ref{sec:Scattering-theory-of} 
we review
the theory and discuss the practical implementation of the method,
which more closely follows the ideas of Ref.\ \onlinecite{Meidan2010},
and differs from that of Ref.\ \onlinecite{Fulga2012c} at some
minor points. The relation to the band-structure-based approach
is discussed in Sec.\ \ref{sec:3}. In section \ref{sec:PhaseDiagram}
we present the phase diagram in the mass--disorder strength plane. In
contrast to Ref.\ \onlinecite{Kobayashi2013} we see no evidence of a ``defeated
WTI'' phase. We conclude in section \ref{sec:Conclusion}. Two
appendices contain details on analytic modeling of the scattering matrix
for the clean limit and an assessment of finite-size effects.

\section{Scattering theory of three-dimensional topological insulators\label{sec:Scattering-theory-of}}

\begin{figure}
\centering{}\includegraphics{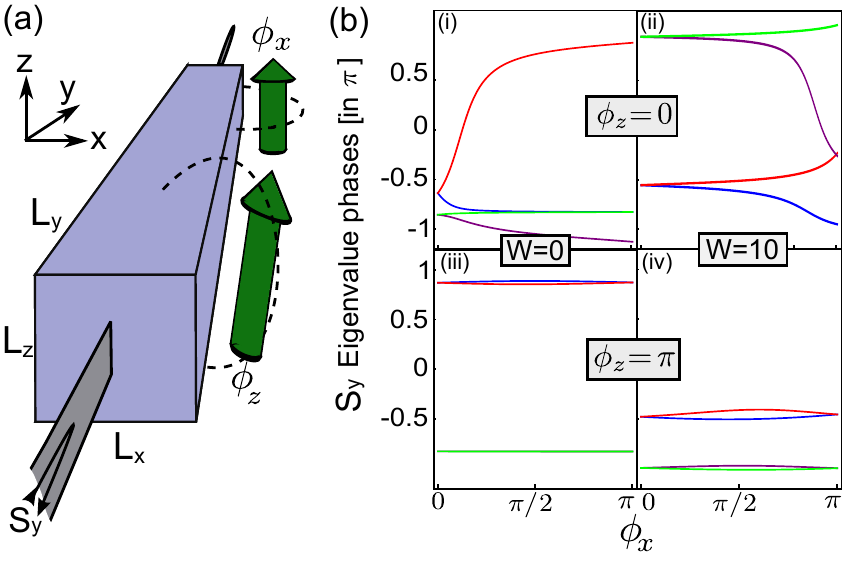}
\caption{\label{fig:lead and flux}(Color online) (a) Setup of the scattering
problem with leads in $y$-direction and twisted periodic boundary
conditions in $x$- and $z-$directions. (b) Typical eigenphase
evolution of $S_{y}(\phi_{x},\phi_{z})$ under continuous variation
$\phi_{x}:0\rightarrow\pi$ for $m_{0}=-1$ red and $\mu=0$ in the clean case $W=0$
{[}panels (i), (iii){]} and with potential disorder $W=10$
{[}panels (ii), (iv){]}. For $\phi_{z}=0$ {[}panels (i), (ii){]}
a nontrivial winding is obtained, while $\phi_{z}=\pi$
{[}panels (iii), (iv){]} shows a trivial winding.} 
\end{figure}

The tight-binding model we consider is a variant of the widely-used 
low-energy effective Hamiltonian of the
$Bi_{2}Se_{3}$ material family.\cite{Zhang2009,Liu2010,Imura2012} In the absence
of disorder the momentum-representation Hamiltonian reads
\begin{eqnarray}
H_{0}(\mathbf{k})&=&
  \tau_{z}\left[m_{0}+2m_{2}\sum_{i=x,y,z}\left(1-\cos k_{i}\right) \right]
  \nonumber \\ && \mbox{}
  +A\tau_{x}\sum_{i=x,y,z}\sigma_{i}\sin k_{i}+\mu,\label{eq:H_clean}
\end{eqnarray}
where Pauli matrices $\sigma_{i}$ and $\tau_{i}$ refer to spin-
and orbital degrees of freedom, respectively. For definiteness, we
set $A=2m_{2}$ and choose energy units such that $m_{2}=1$. 
The system has time reversal symmetry, 
  $TH_{0}(\mathbf{k})T^{-1}=H_{0}(-\mathbf{k})$,
inversion symmetry,
  $IH_{0}(\mathbf{k})I^{-1}=H_{0}(-\mathbf{k})$,
and, if $\mu=0$, particle-hole symmetry
  $PH_{0}(\mathbf{k})P^{-1}=-H_{0}(-\mathbf{k})$. Here $T=i\sigma_{y}K$ is the time-reversal operator 
($K$ complex conjugation, $T^{2}=-1$), $I=\tau_{z}$ the inversion operator,
and $P=\tau_{y}\sigma_{y}K$ the particle-hole conjugation operator 
($P^{2}=1$). 

The full Hamiltonian
\begin{equation}
  H = H_0 + V \label{eq:H}
\end{equation}
includes an on-site disorder potential $V$ that respects time 
reversal symmetry. The most general form of the disorder potential $V$ is
\begin{eqnarray}
  V(\vr) & = & \sum_{\mathbf{r}}\sum_{d=1}^{6}w_{d,\mathbf{r}}\left(\sigma\tau\right)_{d},\label{eq:V}
\end{eqnarray}
where the summation is over all lattice sites $\vr$, and $\left\{ \sigma\tau\right\} =\left\{ 1,\tau_{x},\tau_{y}\sigma_{x},\tau_{y}\sigma_{y},\tau_{y}\sigma_{z},\tau_{z}\right\} $. The amplitudes $w_{d,\mathbf{r}}$ are drawn from a uniform distribution in the interval $-W_{d}/2 < w_{d,\vr} < W_{d}/2$. The disorder potential breaks inversion symmetry; the terms $w_1$, $w_3$, $w_4$, and $w_5$ also break particle-hole symmetry. We consider a lattice of size $L_x \times L_{y}\times L_{z}$ and apply periodic boundary conditions in the $x$ and $z$ directions, but open boundary conditions at the surfaces at $y=0$ and $y = L_{y}-1$. Below, we first discuss the case of potential disorder only ($W_{1}\equiv W$, $W_{d} = 0$ for $d > 1$), and return to the other disorder types at the end of our discussion. 

Our main focus will be on the case $\mu=0$ where, without disorder, three different topological phases appear inside the parameter range $m_{0}\in[-5,4]$, which is the parameter range we consider here. For $m_{0}<-4$ the model is in the WTI phase, with topological indices $(\nu_0,\nu_{x}\nu_{y}\nu_{z})=(0,111)$; for $-4<m_{0}<0$ it is in the STI phase with indices $(1,000)$; for $m_0 > 0$ the system is in the OI phase with indices $(0,000)$. The inversion symmetry of the clean model with $\mu=0$ ensures that bulk gap closings exist at the topological phase transitions at $m_{0}=0$ and $-4$ only.\cite{Murakami2007}

In order to obtain a scattering matrix, we open up the system
by attaching two semi-infinite, translation- and time-reversal 
invariant leads to both surfaces orthogonal to, say, the $y$-direction,
as shown in Fig. \ref{fig:lead and flux}(a). The leads are described by
a tight-binding model, defined on the same lattice grid as the bulk
insulator. In principle, for the scattering matrix method, the leads can be generic and are to be chosen as simple as possible for fast computation. However, for reasons related to numerical robustness, we choose a lead that is one site wide in the $x$ direction, but two sites wide in the $z$ direction.
(We refer to Appendix\ \ref{app:a} for a detailed discussion why in this case
a strictly one-dimensional chain is less well suited for the purpose 
of topological classification.) 
The $y$ coordinates of the lead sites $\vr$ are $y < 0$ and $y \ge L_y$.
Without loss of generality, the $x$ and $z$ coordinates of the lead sites
are fixed at $x=0$ and $z=0$, $1$. Using $\ve_x$ and $\ve_z$ to denote
unit vectors in the $x$ and $z$ directions, respectively, the 
Hamiltonian for the left lead reads (see also Appendix\ \ref{app:a})
\begin{eqnarray}
  H_{\mathrm{L}} & = & \sum_{y<0}
  \sum_{z=0,1} 
  \left[\vphantom{M_M^M}
  t_{0}
  |\vr\rangle \left(\tau_{y}\sigma_{y}+\tau_{y}\sigma_{z}+\mu\right)
  \langle \vr |
  \label{eq:HL}
  \right. \nonumber \\ && \left. \mbox{}
  + 
  i t_y \left( |\vr \rangle \tau_x \sigma_x \langle \vr - \ve_y| -
  |\vr - \ve_y \rangle \tau_x \sigma_x \langle \vr | \right)
  \right. \nonumber \\ && \left. \mbox{} 
  + \delta_{z,0} t_z
  (|\vr \rangle \langle \vr + \ve_z| + |\vr + \ve_z \rangle \langle \vr |)
  \vphantom{M_M^M}
  \right]
\end{eqnarray}
with lattice vector $\vr = (0,y,z)$. 
In our calculations we have set $t_{0}=t_{z}=1$ and $t_{y}=2/5$. For
this choice of parameters the lead supports four right-propagating modes 
and their left-propagating time reversed partners. The coupling between 
the leads and the
bulk sample is described by the coupling term
\begin{equation}
  W_{\rm L} = i \gamma t_y \sum_{z=0,1} 
  ( | \vr \rangle \tau_x\sigma_x \langle \vr - \ve_y| -
  |\vr - \ve_y \rangle \tau_x\sigma_x \langle \vr |),
\end{equation}
with $\vr = (0,0,z)$.
Similar expressions apply to the Hamiltonian $H_{\rm R}$
of the right lead and the coupling $W_{\rm R}$ between the right lead and
the sample. In our calculations we have chosen the value $\gamma = 5$,
optimized empirically for the numerical detection of the scattering 
resonances.

To find the topological invariants for a disordered sample, we employ 
the twisted boundary conditions method.\cite{Niu1985,Leung2012}
This amounts to inserting additional phase factors $e^{i\phi_{x}}$ and
$e^{i \phi_{z}}$ in 
the hopping matrix elements connecting sites at $x=0$ and $x=L_{x}-1$, and
$z = 0$ and $z=L_{z}-1$, respectively. The resulting system can be 
thought of as a large unit cell defined on a torus with two 
independent Aharonov-Bohm fluxes threading the two holes around the $x$
and $z$ axes. For the purpose of classifying insulating phases it is
sufficient to focus on the reflection matrix $S_{y}(\phi_{x},\phi_{z})$ 
of the left 
($y<0$) lead, which is a unitary matrix for an insulating sample. 
For our choice of parameters, the leads have four propagating modes at the Fermi energy ($\varepsilon=0$),
so that $S_y$ is a $4 \times 4$ matrix.

To obtain topological invariants from the scattering matrix, we note
that, because of time-reversal invariance, $S_{y}$ satisfies the
condition
\begin{equation}
S_{y}(\phi_{x},\phi_{z})V=-V^{\rm T}S_{y}^{\rm T}(-\phi_{x},-\phi_{z})\label{eq:S_TRS}
\end{equation}
where $\rm{T}$ denotes the matrix transpose and the unitary matrix $V$ describes the action of the time reversal
operator $T$ in the space of scattering states.\cite{Fulga2012c}
Since $T$ flips the sign of the velocity $v=dE/dk$, it connects
incoming and outgoing modes, $T\psi_{n}^{\rm in}=\sum_{k}V_{nk}
\psi_{k}^{\rm out}$. Reference \onlinecite{Fulga2012c} chooses a
convention wherein, after redefinition of the incoming scattering
states, $S_{y}V\rightarrow S_{y}^{\prime}$ the scattering matrix 
becomes antisymmetric at the ``time-reversal invariant fluxes'' 
$\phi_{x,z} = 0,\pi$, and, thus, acquires the same symmetry properties as 
the matrix $w(\mathbf{k})$ used by Fu and Kane to classify time-reversal
invariant topological insulators without disorder in terms of their band 
structure.\cite{Fu2007a} 
Here, we follow the formulation of scattering theory as it is most
commonly used in the theory of quantum 
transport,\cite{Beenakker1997,NazarovBlanterBook}
in which one makes the choice $V V^{*}=-1$. At the time-reversal
invariant fluxes $\phi_{x,z} = 0,\pi$ this gives the condition that $S_y$ 
is ``self dual'', $S_y^{\rm{T}}=V^{-1} S_y V$. Then
Kramers degeneracy ensures that the eigenphases $\{ e^{i \theta_j}\}_{j=1,...,4}$ of
$S_y$ are twofold degenerate at $\phi_{x,z} = 0, \pi$. 
The topological classification rests on the eigenvalue evolution as 
one of the fluxes $\phi_x$ or $\phi_z$ changes from $0$ to $\pi$,
so that the system evolves from one time-reversal invariant flux
configuration into another one:\cite{Meidan2011}
In the topologically trivial case, labeled by $\mathcal{Q}[S_{y}]=0$,
degenerate eigenvalue pairs, which generically split upon departing 
from a time-reversal invariant flux, are reunited upon reaching the 
other time-reversal invariant fluxes. In the nontrivial case, which
we label by $\mathcal{Q}[S_{y}]=1$, the eigenphases from a degenerate
pair are united with eigenphases from different pairs. (If $S_y$ is
a $2 \times 2$ matrix, so that there is only a single eigenvalue pair,
the question of topological triviality is connected to the winding
of the eigenphase pair around the unit circle.\cite{Meidan2011}) One easily 
verifies that this definition is independent of the choice which 
eigenphase pair is being ``tracked'': if one eigenphase pair 
``switches partners'', then all eigenphase pairs must do so. Similar
considerations have been applied to Kramers degenerate energy level
pairs in order to argue for topological non-triviality of time-reversal
invariant topological 
insulators.\cite{Fu2006,Fu2007a,Nomura2007} The strong and 
weak topological invariants of the sample are then defined 
as\cite{Fulga2012c}
\begin{eqnarray}
\nu_{0} & = & \{\mathcal{Q}\left[S_{y}\left(\phi_{z}=0,\phi_{x}:0\rightarrow\pi\right)\right] \nonumber \\
 & & \mbox{} + \mathcal{Q}\left[S_{y}\left(\phi_{z}=\pi,\phi_{x}:0\rightarrow\pi\right)\right]\}\, 
\mathrm{mod}\, 2 \nonumber \\
  &=& \{\mathcal{Q}\left[S_{y}\left(\phi_{x}=0,\phi_{z}:0\rightarrow\pi\right)\right] \nonumber \\ && \mbox{} + \mathcal{Q}\left[S_{y}\left(\phi_{x}=\pi,\phi_{z}:0\rightarrow\pi\right)\right]\}\, 
\mathrm{mod}\, 2, \label{eq:nu0} \\
\nu_{x} &=& \mathcal{Q}\left[S_{y}\left(\phi_{x}=\pi,\phi_{z}:0\rightarrow\pi\right)\right], \label{eq:nux} \\
\nu_{z} & = & \mathcal{Q}\left[S_{y}\left(\phi_{z}=\pi,\phi_{x}:0\rightarrow\pi\right)\right].\label{eq:nuz}
\end{eqnarray}
The two expressions for $\nu_{0}$ are equivalent, because the evolution
of an eigenphase pair for a contractable loop in the $\phi_{x}$, 
$\phi_{z}$-plane is always trivial.
The relations (\ref{eq:nu0})--(\ref{eq:nuz}) 
remain valid under circular permutation of spatial
indices, so that, \textit{e.g.},
the weak topological index $\nu_{y}$ can be calculated by attaching a 
lead in the $x$ or $z$ directions.

\begin{figure}
\centering{}\includegraphics[scale=0.7]{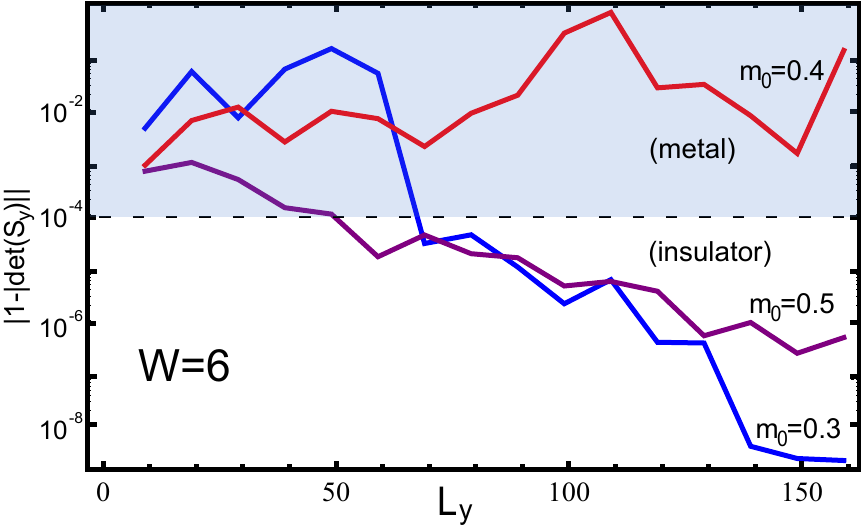}
\caption{\label{fig:absDetS}(Color online) 
Evolution of $|\det[S_{y}(\phi_{x,z}=0)]|$
as a function of $L_{y}$ for $L_{x,z} = 11$ and 
a specific disorder realization with disorder strength $W=6$ and $\mu=0$.
The three curves are for $m_0$ are $0.3$, $0.4$, and $0.5$, 
corresponding to the STI
phase, the immediate vicinity of the topological phase transition, and
the OI phase, respectively. The dashed line indicates the empirical
cut-off used in the calculations.}
\end{figure}

Using the \noun{Kwant} software package,\cite{Groth2013} we performed numerical calculations of $\nu_{0}$ and $\nu_{z}$
on a system with dimensions $L_{x,z}\simeq9$ and
variable $L_{y}=9..160$. Here the length $L_{y}$ was increased until an
(almost) unitary reflection matrix $S_y(\phi_x,\phi_z)$ was found, 
where we used the condition $||\mathrm{det}S_{y}|-1|<10^{-4}$ as an
empirical cut-off where unitarity is reached. The possibility of 
large system sizes $L_{y}$ is needed to accommodate cases with a 
long localization length, as it occurs close to a topological phase 
transition. If the condition $||\mathrm{det}S_{y}|-1|<10^{-4}$ could
not be met for $L_{y}\leq160$ the system is empirically labeled as 
metallic. (Note that a full assessment of the metal/insulator
transition requires an analysis of the scaling behavior of 
conductivity, which is beyond the scope of this work.) The approach
to a unitary scattering matrix is illustrated in Fig.\
\ref{fig:absDetS}, which shows the evolution of 
$|\det[S_{y}(\phi_{x,z}=0)]|$ as a function of $L_{y}$ at disorder 
strength $W=6$ across the OI-STI transition for three different 
values of $m_{0}$.
During the sweep of the flux $\phi_{x}$ and $\phi_{z}$, the eigenphases
have been tracked using a dynamical step-width control, allowing to 
resolve sharp features in the eigenphase trajectory. 
Note that the use of twisted boundary conditions in the $x$ and $z$
directions allows us to chose moderate $L_{x,z}$, since we are not 
required to separate any surface states. We found the system size 
$L_{x}$, $L_{z} = 9$ 
sufficient to suppress finite-size issues: Beyond a parity effect
for $L_{x,z}$ (see the discussion in the next section) there is no dependence of the results on increased $L_{x,z}$, see Appendix \ref{app:b}.

As an example, Fig. \ref{fig:lead and flux}(b) shows a typical eigenphase
evolution for $m_{0}=-1$ and $\mu=0$ in the clean and a disordered case
($W=0$ and $W=10$). For $\phi_{z}=0$ a topologically nontrivial winding
is obtained, while $\phi_{z}=\pi$ shows a trivial winding for both
the clean and the disordered case. With Eqs. \eqref{eq:nu0} and \eqref{eq:nuz}
we obtain $\nu_{0}=1$ and $\nu_{z}=0$, respectively.
Similarly, we confirmed $\nu_{x}=\nu_{y}=0$ which, in summary, leads
to $(\nu_0,\nu_x\nu_y\nu_z)=(1,000)$ for the particular points in
parameter space. 

\section{Comparison with band-structure-based approach}\label{sec:3}

In this section, we focus on $\mu=0$. For a clean bulk system, the topological indices $\nu_0$ and $\nu_{x,y,z}$
can also be calculated from the band structure. The weak indices one 
obtains from the scattering approach agree with those for the bulk system 
if and only if the sample dimensions $L_{x}$, $L_{y}$, and $L_{z}$ are 
odd. (For even sample dimension, the scattering method yields trivial weak indices.) The advantage of the scattering approach is that the
weak indices can be calculated for a disordered system as well.

In order to show that the scattering-matrix-based topological indices
of Eqs.\ (\ref{eq:nu0})--(\ref{eq:nuz}) are the same as the 
band-structure based indices if the sample dimensions are odd, 
we make use of the relation between
scattering phases and bound (surface) states: 
A surface state exists at energy
$\varepsilon$ if and only if $S_y$ for energy $\varepsilon$ has an eigenphase
$\pi$. This relation follows from the observation that capping the 
lead by a ``hard wall'', which has scattering matrix $-1$, restores 
the original surface state spectrum without coupling to an external lead.
A nontrivial winding requires that an \textit{odd} number of eigenphases
passes the reference phase $\pi$ upon sweeping the fluxes
$\phi_x$ and $\phi_z$ as specified in Eq.\ 
(\ref{eq:nu0})--(\ref{eq:nuz}), whereas an even number of eigenphases
passes the reference phase $\pi$ if the winding is 
trivial.\cite{Meidan2011} Note, that depending on the definition of the lead modes, the numerical value of the reference phase might differ from $\pi$.
(In Appendix \ref{app:a},
we show that for the clean and weak coupling limit all phase winding signatures
can be reproduced quantitatively
from an analytical calculation of the scattering matrix in terms of the surface
states at the $y=0$ surface.)

\begin{figure}
\centering{}\includegraphics[scale=1]{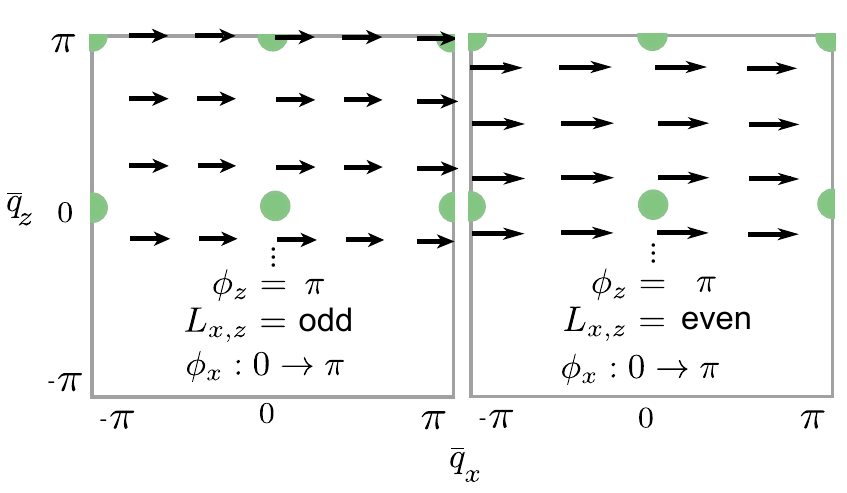}
\caption{\label{fig:4_types}(Color online)
Brillouin zone for a surface orthogonal to the $y$ direction. Black arrows indicate the trajectories of surface wave-vectors $\bar{\mathbf{q}}=(q_x,q_z)$ corresponding
to $L_x$ and $L_z$ both odd (left) or even (right) for fixed $\phi_z=\pi$ and a sweep of $\phi_x$ from $0$ to $\pi$. Dots indicate time-reversal-invariant momenta which are possible positions for surface Dirac cones at $\mu=0$.}
\end{figure}

In a clean system, translation invariance in the $x$ and $z$ directions
implies that the surface states are labeled by a wave-vector
$\bar{\vq} = (q_x,q_z)$ in the surface Brillouin zone. Possible Dirac cones in the 
$(q_x,q_z)$ plane are centered around the four time-reversal-invariant
momenta $(q_x,q_z) = (0,0)$, $(0,\pi)$, $(\pi,0)$, and $(\pi,\pi)$, see Fig. \ref{fig:4_types}.
For a finite-size sample with twisted boundary conditions, only 
discrete values $q_x = (2 \pi n - \phi_x)/L_x$, $q_z = (2 \pi n - 
\phi_z)/L_z$ are allowed. A resonance (\textit{i.e.} scattering phase $\pi$) is found if 
one of the allowed
$\bar{\vq}$ vectors crosses one of the surface Dirac cones.

For definiteness, we now consider the weak index $\nu_z$, which
is determined by the phase winding $\mathcal{Q}$
along the path $\phi_x:0 \to \pi$
at fixed $\phi_z = \pi$. While sweeping $\phi_x$, the allowed
$\bar{\vq}$ values build a set of trajectories in the $(q_x,q_z)$
plane, which are shown in Fig.\ \ref{fig:4_types} for the cases 
of $L_x$ and $L_z$ even or odd. From inspection of Fig.\ \ref{fig:4_types}
one immediately concludes, that a 
Dirac cone gives rise to an odd number of scattering
resonances if and only if its center is at one of the ``allowed''
$\bar{\vq}$ vectors for $\phi_x=0$ or for $\phi_x=\pi$,
which requires odd $L_z$ for Dirac points with $q_z=\pi$.
Hence,
we conclude that if and only if $L_z$ is odd, the index 
$\nu_z$ of Eq.\ (\ref{eq:nuz})
measures the parity of the number of Dirac points with $q_z = \pi$.
Similarly, the index $\nu_x$ of (\ref{eq:nux})
measures the parity of the number of Dirac points
with $q_x = \pi$ if and only if $L_x$ is odd, whereas the index $\nu_0$ 
of Eq.\ (\ref{eq:nu0})
measures the parity of the total number of Dirac points for both
even and odd sample dimensions. 
In all three cases, the parities of number of Dirac points corresponds
to the very same quantities as those that are
computed from the band structure.\cite{Fu2007,Hasan2010,Bernevig2013}

There is a simple argument that shows that the scattering-matrix-based
weak indices are always trivial if the sample dimensions are even, 
irrespective of the value of the bulk index: 
Any three-dimensional weak topological insulator is adiabatically connected to a stack of two dimensional topological insulators. The stacking direction can be taken to be $G_\nu=(\nu_{x},\nu_{y},\nu_{z})$. A ``mass term'' that 
couples these layers in pairs connects the system adiabatically to a trivial insulator.\cite{Ringel2012,Mong2012,Liu2012} If $L_x$,  $L_y$, $L_z$ are all even, such a mass term can be applied for any $G_\nu$. Since the indices of Eqs.\ (\ref{eq:nux}) and
(\ref{eq:nuz}) are true topological invariants, they cannot change
upon inclusion of such a mass term, \textit{i.e.}, they can only 
acquire a value compatible with the topologically trivial phase.

For odd sample dimensions this argument does not apply and, as is
shown above, for the clean case, the topological indices derived from the
scattering matrix agree with the indices obtained from the band
structure.

\section{Phase diagram in the presence of disorder}\label{sec:PhaseDiagram}

\begin{figure}
\begin{centering}
\includegraphics{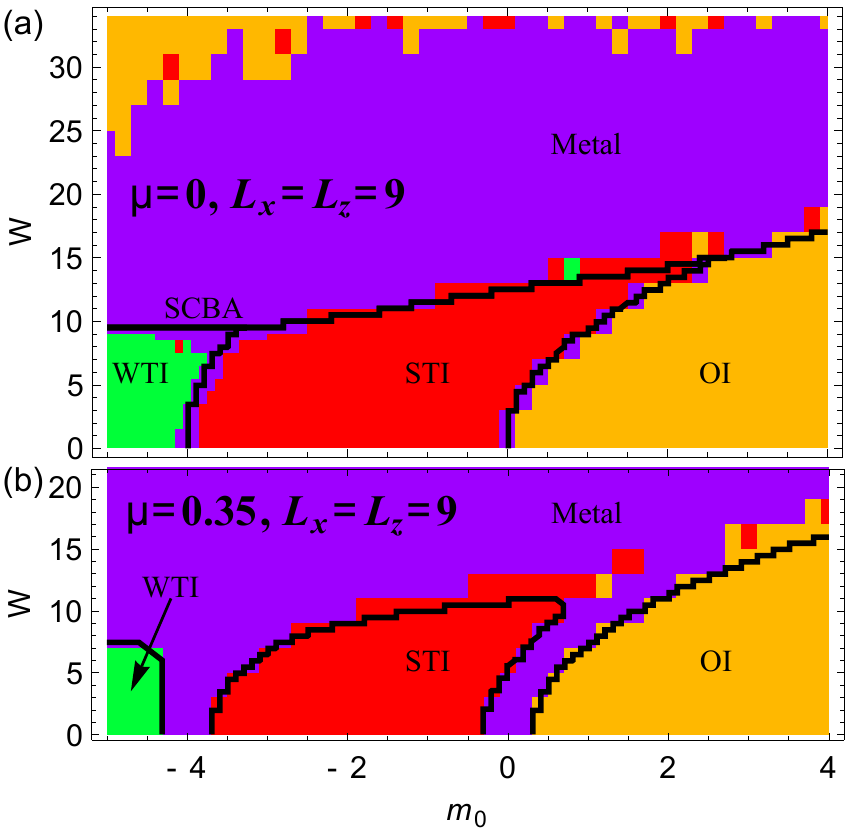}
\par\end{centering}

\caption{\label{fig:phase diagram}(Color online) Topological phase diagram
of model $H$ as calculated with the scattering matrix method with
potential disorder in the mass ($m_{0}$) -- disorder strength ($W$) plane for $\mu=0$ (a) and $\mu=0.35$ (b). The sample dimensions are $L_{x,z}=9$ $L_{y}\leq160 $. Solid lines denote
the SCBA phase boundaries of insulating phases with $\mathrm{Im}\Sigma=0$.}
\end{figure}

We now discuss the $\mathbb{Z}_{2}$ phase diagram of the three-dimensional
Hamiltonian $H$ in the $(m_{0},W)$ parameter plane with
potential disorder. We study the cases $\mu=0$ and $\mu=0.35$. Topological indices $\nu_{0}$ and $\nu_{z}$
are computed as described in Sec.\ \ref{sec:Scattering-theory-of}
for a dense grid of parameter values. The result is shown in Fig.
\ref{fig:phase diagram}. For $\mu=0$, it confirms similar topological phase diagrams
computed on the basis of conductance and scaling methods as in Refs.\
\onlinecite{Ryu2012,Kobayashi2013}. Due to the large maximum system size
of \textcolor{red}{${\color{black}9\times160\times9}$} we relied
on self averaging and worked with only a single disorder realization
per point in parameter space. The results indicate that this is indeed
justified for the range of weak and moderate disorder strengths; only
for the strong disorder region $W>25$, where Anderson localization
and a trivial insulator is expected, a minority of data points yields
diverging results. 

Studies of disorder effects of the three-dimensional quantum critical point between STI and OI at $\mu=0$ employing the
self consistent Born approximation\cite{Shindou2009}, renormalization
group\cite{Goswami2011} or a numerical approach\cite{Kobayashi2014} show the existence of a critical disorder
strength below which a direct phase transition without extended metallic
phase is realized. This conclusion however is valid only for systems
with chemical potential at the clean band-touching energy (here $\mu=0$) which also
preserve inversion symmetry (after disorder average). Indeed, our
numerical results for $\mu=0$ show that the width of the metal region at the $m_{0}$-induced
transition between WTI, STI and OI, for weak disorder is considerably
smaller than in other studies of the $\mathbb{Z}_{2}$ invariant for
disordered systems,\cite{Leung2012,Fulga2012c} indicating that finite-size
effects are much less severe for the large system sizes we can reach.
Further indication for the successful suppression of finite-size effects
is that the phase diagram in Fig. \ref{fig:phase diagram}(a) remains
unchanged if we increase the system volume by $50\%$ to $11\times160\times11$,
see Appendix \ref{app:b}.)

An analytical approach to disordered topological
insulators is the calculation of the disorder averaged self-energy
$\Sigma$ using the self-consistent Born approximation 
(SCBA).\cite{Shindou2009,Guo2010a}
Due to symmetry arguments, $\Sigma$ can be expanded as 
$\Sigma_{z}\tau_{z}+\Sigma_{0} \tau_0$, where $\tau_0$ is the $2 \times 2$
unit matrix, and the SCBA equation reads \cite{Guo2010a,Shindou2009}
\begin{equation}
  \Sigma=\sum_{d=1}^{6}
  \frac{W_{d}^{2}}{12}\sum_{\mathbf{k}\in BZ} (\sigma\tau)_{d}\frac{1}{i\delta-H_{0}(\mathbf{k})-\Sigma} (\sigma\tau)_{d},\label{eq:SCBA}
\end{equation}
where the notation $(\sigma\tau)_{d}$ was introduced below Eq.\ (\ref{eq:V}).
Consequently, the disorder averaged propagator features renormalized
mass and chemical potential values $\bar{m}=m_{0}+\mathrm{Re}\Sigma_{z}$
and $\bar{\mu}=\mu-\mathrm{Re}\Sigma_{0}$, respectively.  If $\mathrm{Im}\Sigma=0$ and $\bar{\mu}$ in the bands above and below energies $\pm\mathrm{min}(|\bar{m}|, |\bar{m}+4|)$ the system is metallic; otherwise, if $\bar{\mu}$ is in the bandgap, the value of $\bar{m}$ determines the nature of the resulting insulator:
For $0<\bar{m}$ we expect an OI, $-4<\bar{m}$ yields a WTI and $-4<\bar{m}<0$
indicates a STI. Nonzero
imaginary parts, $\mathrm{Im}\Sigma_{z}$ and $\mathrm{Im}\Sigma_{0}$
translate into a finite lifetime $\tau<\infty$ and a finite density
of states at the Fermi level, indicating either a compressible diffusive
metal phase\cite{Goswami2011} or, if these states are localized, an insulator. SCBA cannot distinguish between both possibilities.

The coupled set of SCBA equations \eqref{eq:SCBA} is numerically
solved self-consistently. For potential disorder ($W_d=0$ for $d>1$), the resulting
phase boundaries of insulating phases with $\mathrm{Im}\Sigma=0$ are shown in Fig. \ref{fig:phase diagram} as solid
lines. For $\mu=0$, we find excellent agreement of the SCBA phase boundaries with
the results from the scattering matrix method. Since SCBA as a disorder-averaged theory is free
of finite size effects, this further supports the applicability
of the scattering matrix results in the thermodynamic limit. The situation is different for $\mu=0.35$, where for strong
disorder ($W \gtrsim 10$) the insulating states slightly but numerically significantly exceed the regions where $\mathrm{Im}\Sigma=0$ as obtained from SCBA, indicating localized states at the Fermi energy. A similar observation was reported in Ref. \onlinecite{Leung2012}.

In closing, we comment on the effect of the five remaining disorder types.
By inspection of Eq. \eqref{eq:SCBA} we find that mass-type disorder,
$(\sigma\tau)_{6}=\tau_{z}$, has the same effect as pure 
potential disorder, \textit{i.e.}, bending the phase boundaries between insulating phases to increased
values of $m_{0}$. All other disorder types have the opposite effect
on $\bar{m}$, as was noticed for the two dimensional case in Ref.\ \onlinecite{Song2012}. We have confirmed the agreement between scattering
matrix results and the trends predicted by SCBA in these cases (results
not shown). We conclude that qualitative features of the phase diagram,
like, for example, the occurrence of a disorder-induced 
topological Anderson insulator transition,
crucially rely on the microscopic details of the disorder potential. 

\section{Conclusion \label{sec:Conclusion}}

We have demonstrated the potential of the scattering matrix method
for the computation of $\mathbb{Z}_{2}$ topological indices for a
three-dimensional disordered tight-binding model featuring strong
and weak topological phases. We studied the $\mathbb{Z}_{2}$
phase diagram in the mass - disorder plane for system sizes up to $11\times160\times11$ and found excellent agreement with SCBA predictions.
The latter have been studied in the literature
before\cite{Shindou2009,Goswami2011,Ryu2012} (only for the OI/STI case and for $\mu=0$) but have never been compared quantitatively
to a real-space disordered three-dimensional TI tight-binding model. We conclude that SCBA should have predictive value also for similar
scenarios. In particular, we showed that SCBA is quantitatively correct also for finite chemical potential and weak disorder, where extended metal regions occur even for weak disorder, whenever the (renormalized) chemical potential lies within a bulk band. This possibility has been overlooked in Ref. \onlinecite{Yamakage2013a}. For the insulator-metal transition at larger disorder strength, SCBA's precision suffers from its inherent inability to take into account localization effects\cite{Yamakage2013a} which occur at the edges of topological nontrivial bands.

The scattering matrix method can be regarded as complementary to a finite-size scaling analysis. While the latter is ideally suited to detect a phase boundary, the scattering matrix method can unambiguously identify the topological
phase at each parameter point where the system is insulating.  This proves the nontrivial $\mathbb{Z}_{2}$
nature of the TAI phase without referring to adiabatic connection
to the clean STI phase or involving other indirect arguments. For the disordered WTI, we find no evidence for a ``defeated WTI''
region in the phase diagram, as suggested recently in Ref.\ \onlinecite{Kobayashi2013}.
We point out that the scattering matrix method should be an ideal
tool to identify the topological invariants for (so far hypothetical) disordered topological
phases that are not adiabatically connected to the clean case.

The scattering matrix method is able to find weak indices even if
the strong index is nonzero, as has been checked using a modified Hamiltonian
$H$ (as in Ref.\ \onlinecite{Imura2012}) with anisotropic mass parameters
which realizes many more topological phases, e.g. $(\nu_0,\nu_x\nu_y\nu_z)=(1,001)$.
Moreover, our results explicitly demonstrate the intricate interplay
between system size and topological phase in the parameter region
supporting a WTI phase. Adding a single layer to the system can change
the topological phase from OI to WTI or vice versa, a behavior not
reflected in conductance simulations. The case of a disordered WTI phase has been previously discussed in Refs. \onlinecite{Fu2012, FulgaArxiv2012}, where it is argued that average translational symmetry in stacking direction is sufficient to protect the weak topological insulator phase. This is in agreement with our findings since an odd number of stacked layers prohibits any average translational symmetry breaking while such a dimerization can be adiabatically applied to an even number of layers.

\acknowledgements

We thank A. Akhmerov, C. Groth, X. Waintal and M. Wimmer for making the \noun{Kwant} software package available before publication.  We thank a thoughtful referee for pointing out the problem with SCBA in Ref. \onlinecite{Yamakage2013a}. Financial support was granted by
the Helmholtz Virtual Institute ``New states of matter and their
excitations'' and by the Alexander von Humboldt Foundation in the framework of the Alexander von Humboldt Professorship, endowed by the Federal Ministry of Education and Research.

\appendix

\section{Analytic modeling of the phase winding in the clean limit}
\label{app:a}

In the clean case, it is possible to understand the scattering
matrix eigenvalue phase winding signatures (and thus the topological
classification) from a microscopic point of view. 
We employ the Fisher-Lee relation\cite{Fisher1981} to calculate the elements of the scattering matrix from the retarded Green function $G^R$,
\begin{equation}
S_{nm}=-\frac{\sqrt{v_{n}}}{\sqrt{v_{m}}} 1_{nm}+i\sqrt{v_{m}}\sqrt{v_{n}}G_{nm}^{R} \label{eq:FisherLee}
\end{equation}
where the right hand side represents the current in outgoing lead mode $n$ after a normalized local excitation of incoming mode $m$. The mode velocities $v_n$ and $v_m$ link this quantity to the usual amplitude propagation described by $G^R$ and any direct transition into outgoing modes ($\propto 1_{nm}$, not contributing to the system's scattering matrix) is subtracted.
The Green function depends on the scattering region (\textit{i.e.} the topological insulator surface), the lead and their mutual coupling.
We first discuss the effective description of the topological insulator surface and specify a simplified lead $H_{\mathrm{L}}^\prime$. We then compare the analytical prediction with the full-scale numerical calculation. Finally we motivate the lead choice in the main text, $H_{\mathrm{L}}$.

\subsection{Surface states and surface Hamiltonian}

Following the convention of the main text, we consider a clean topological
insulator described by Eq.\ (\ref{eq:H_clean}), occupying the half 
space $y \ge 0$. We make the same parameter choice as described in 
Sec.\ \ref{sec:Scattering-theory-of}. For energies in the bulk gap, a description in terms of the effective surface theory is sufficient. The Bloch wavefunctions
for the surface states at surface momentum $\bar{\mathbf{q}}
= (q_x,q_z)$
close to a Dirac point at momentum $\bar{\mathbf{Q}} =
(Q_x,Q_y)$
can be found using the method applied in Ref. \onlinecite{Liu2012}. For
the STI ($-4<m_{0}<0$) the two surface states around the single Dirac point at 
$\bar{\mathbf{Q}}=\left(0,0\right)$ read
\begin{eqnarray}
\psi_{\bar{\mathbf{q}}}^{(1)}(x,y,z) & = & \frac{1}{\sqrt{L_x L_z}} e^{i \bar{\vq}\cdot \bar{\vr}}\left(\begin{array}{c}
1/\sqrt{2}\\
0\\
0\\
1/\sqrt{2}
\end{array}\right)\varphi(y),\label{eq:STI_surface_state_spin_up_around-0-1}\\
\psi_{\bar{\mathbf{q}}}^{(2)}(x,y,z) & = & \frac{1}{\sqrt{L_x L_z}} e^{i \bar{\vq}\cdot \bar{\vr}}\left(\begin{array}{c}
0\\
-1/\sqrt{2}\\
1/\sqrt{2}\\
0
\end{array}\right)\varphi(y),\label{eq:STI_surface_state_spin_down_around-0-1}
\end{eqnarray}
in the same basis as Eq. \eqref{eq:H_clean} and with $\varphi(y)$ a normalized, decaying function for $y \to \infty$.\cite{Liu2012} 
In the basis
of these two Bloch states, the effective surface Hamiltonian becomes
a $2 \times 2$ matrix which reads
\begin{equation}
  \bar{H}_{y}^{\rm STI}(\bar{\vq})=A \left(\begin{array}{cc}
q_{x} & -q_{z}\\
-q_{z} & -q_{x}
\end{array}\right).\label{eq:STI surface}
\end{equation}
The constant $A$ was defined in Eq. \eqref{eq:H_clean}.

For the WTI ($m_{0}<-4$) there are four surface bands, which 
form two Dirac
cones centered around $\bar{\mathbf{Q}}_{1}=(\pi,0)$ and
$\bar{\mathbf{Q}}_{2}=(0,\pi)$. The basis states are the 
same as in Eqs.\ (\ref{eq:STI_surface_state_spin_up_around-0-1}) and (\ref{eq:STI_surface_state_spin_down_around-0-1}), but with surface momenta 
$\bar{\mathbf{q}}_j=(q_{j,x},q_{j,z})$ defined around  $\bar{\mathbf{Q}}_j$ for $j=1,2$, respectively.
We find
\begin{equation}
  \bar{H}_{y}^{\rm WTI}(\bar{\vq}_1,\bar{\vq}_2)=A \left(\begin{array}{cc}
\begin{array}{cc}
-q_{1,x} & -q_{1,z}\\
-q_{1,z} & q_{1,x}
\end{array} & 0\\
0 & \begin{array}{cc}
q_{2,x} & q_{2,z}\\
q_{2,z} & -q_{2,x}
\end{array}
\end{array}\right).\label{eq:WTI surface}
\end{equation}

In a system with finite $L_{x,z}$ and given fluxes $\phi_{x,z}$, a finite subset of surface wave-vectors
are compatible with the twisted boundary conditions, see the discussion
in Sec.\ \ref{sec:3}. During the sweep of the ``flux'' $\phi_{x}$ or 
$\phi_{z}$, the allowed $\bar{\vq}$ values form a set of 
trajectories in the surface Brillouin zone, see Fig. \ref{fig:4_types}. 
For an approximate description of the scattering process, it
is sufficient to further restrict the effective surface Hamiltonian to the
few allowed wave-vectors on trajectories which are closest to the Dirac points.
As we will show momentarily, the arrangement of the trajectories in the surface Brillouin zone 
relative to the locations of the gapless points then determines the phase winding structure.

\subsection{Lead and its self-energy}

\begin{figure}
\centering{}\includegraphics{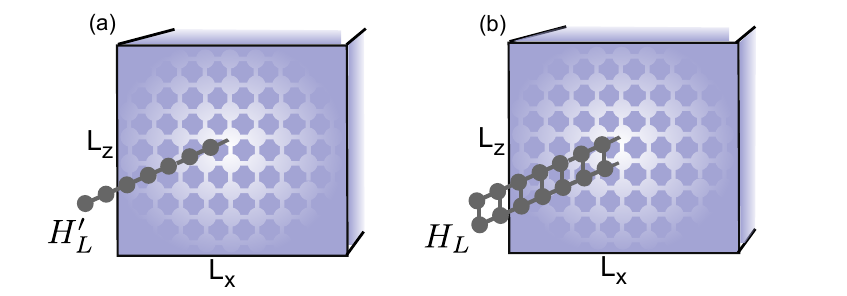}
\caption{\label{fig:RoleOfLead}(Color online) Tight-binding realization of
system with Hamiltonian $H$ with attached lead realized as a translation invariant
chain. In (a), the height of the lead, described by Hamiltonian $H_{\mathrm{L}}^{\prime}$,
is a single lattice site while the lead $H_{\mathrm{L}}$ in (b) has
a height of two lattice sites.}
\end{figure}

The leads are modeled as semi-infinite, translational- and time-reversal invariant tight-binding systems. To motivate the special choice of lead $H_\mathrm{L}$ described
by Eq. \eqref{eq:HL}, we first consider a simpler (thinner) lead as in Fig. \ref{fig:RoleOfLead}(a),
realized as a tight binding chain of lattice sites at coordinates
$|\vr\rangle = (0,y,0)$, with $y < 0$ and Hamiltonian
\begin{equation}
  H_{\mathrm{L}}^{\prime}= \sum_{y < 0} | \vr \rangle H_{\mathrm{hop}}^{\dagger}\langle\vr-\ve_y |
 +| \vr-\ve_y \rangle H_{\mathrm{hop}}\langle\vr |,
  \label{eq:thin lead}
\end{equation}
where $H_{\mathrm{hop}}=t_y\left[\tau_{y}\sigma_{x}-i\tau_{x}\sigma_{y}\right]$. The wavefunctions of the four scattering channels at the four Fermi points $q_y=\pm \pi/4$ and $q_y=\pm 3\pi/4$ are denoted
$| \phi_n^\mathrm{in/out} \rangle$, with $n=1,2,3,4$. They are chosen such that the matrix $V$, defined below Eq. \eqref{eq:S_TRS}, fulfills the condition $V \cdot V^*=-1$.
Finally, the lead $H_\mathrm{L}^\prime$ is coupled to the system $H_\mathrm{S}$ (\textit{i.e.} the topological insulator) by $H_{\mathrm{hop}}$ times a real constant $\gamma$,
\begin{equation}
W^\prime_\mathrm{L}=\gamma \  \lbrack | \vr \rangle H_{\mathrm{hop}}^\dagger \langle\vr-\ve_y |
 +| \vr-\ve_y \rangle H_{\mathrm{hop}}\langle\vr |\rbrack
\end{equation}
for $\vr=(0,0,0)$.

For a semi-infinite lead, the retarded Green function $G^R$ is an infinite dimensional matrix. However, employing the concept of lead self-energy,\cite{Datta1997} the degrees of freedom corresponding to the lead can be eliminated. The calculation of the Green function $G_{nm}^{R}$ in Eq. \eqref{eq:FisherLee} is most efficient if we retain the lead site $y=-1$.
Thus, the lead self-energy should take into account only lead sites $y<-1$. It reads\cite{Datta1997} $\Sigma_{\mathrm{L}}^\prime=H_{\mathrm{hop}}^{\dagger}G_{\mathrm{L}}^\prime H_{\mathrm{hop}}=-it\sqrt{2}$ where $G_{\mathrm{L}}^\prime$ is the Green function of the lead without the coupling $W^\prime_\mathrm{L}$. Finally, at zero energy we have $G^R=(-H_\mathrm{S}-W^\prime_\mathrm{L}-\Sigma_{\mathrm{L}}^\prime)^{-1}$ and 
\begin{equation}
G_{nm}^{R}=\left\langle \phi_{n}^\mathrm{out}(y=-1)\right| G^R \left|\phi_{m}^\mathrm{in}(y=-1)\right\rangle,
\end{equation}
where $\left|\phi_{m}^\mathrm{in}\right\rangle$ and $\left|\phi_{m}^\mathrm{out}\right\rangle$ are incoming and outgoing scattering states for the lead terminated at $y=-1$, \textit{i.e.} without  the coupling $W^\prime_\mathrm{L}$.

\begin{figure}
\centering{}\includegraphics{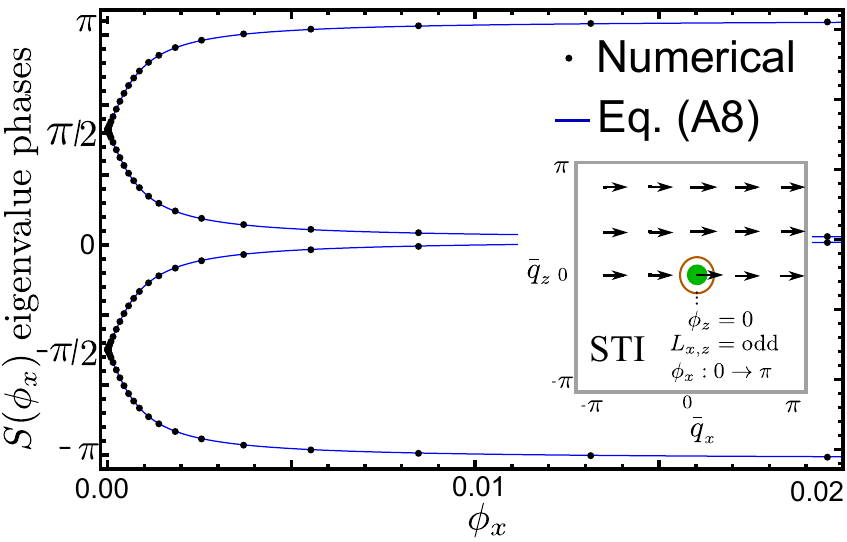}
\caption{\label{fig:phase_winding_plot}
Scattering matrix
eigenvalue phase windings in the case $m_0=-2$ (STI), $\phi_z=0$ with lead as in Eq. \eqref{eq:thin lead} and $L_{x,y,z}=9$ in the weak coupling regime ($\gamma=0.1$, $t_y=1$). The remaining parameters are as in the main text. Dots indicate numerical results based on the full-scale three-dimensional model while solid lines denote analytical results based on Eq. \eqref{eq:SL(x)}. The inset shows the surface Brillouin zone with the position of the Dirac cone for a STI and trajectories of allowed surface momenta for the boundary conditions indicated. The encircled region of the surface Brillouin zone gives rise to the effective model in Eq. \eqref{eq:case_L_setup}.} 
\end{figure}

\subsection{STI phase\label{subsec:STI}}

As a first specific example we consider the case of a strong topological insulator, for which the surface Hamiltonian has a single Dirac cone centered at $\bar \vQ = (0,0)$. We chose $m_0=-2$ since then $\varphi(y)=\delta_{y,0}$, see Ref. \ \onlinecite{Liu2012}. Employing the boundary conditions for, say, $\phi_z=0$ and $L_{x,z}$ odd, the resulting trajectories for the surface momenta are shown in Fig. \ref{fig:phase_winding_plot} (inset). For the effective low energy theory (encircled region in the surface Brillouin zone) we find from Eq. \eqref{eq:STI surface}
\begin{equation}
\bar{H}_{y}^\mathrm{STI}\left(\phi_{x}\right)=A\left(\begin{array}{cc}
\phi_{x}/L_x & 0\\
0 & -\phi_{x}/L_x
\end{array}\right).\label{eq:case_L_setup}
\end{equation}
In order to calculate the Green function $G^{R}$ we assume weak system-lead coupling $\gamma$. Then $H_{\mathrm{S}}$ can be approximated by the ideal effective surface theory without lead, Eq. \eqref{eq:case_L_setup}, and we find in the basis of Eq. \eqref{eq:case_L_setup} and Eq. \eqref{eq:thin lead}
\begin{multline*}
\left(G^{R}\right)^{-1}=\\
\left(\begin{array}{cccccc}
\frac{A}{L_{x}}\phi_{x} & 0 & \frac{(i-1)\gamma}{\sqrt{2L_{x}L_{z}}} & 0 & 0 & \frac{(1-i)\gamma}{\sqrt{2L_{x}L_{z}}}\\
0 & \frac{-A}{L_{x}}\phi_{x} & 0 & \frac{-(1+i)\gamma}{\sqrt{2L_{x}L_{z}}} & \frac{-(1+i)\gamma}{\sqrt{2L_{x}L_{z}}} & 0\\
\frac{-(1+i)\gamma}{\sqrt{2L_{x}L_{z}}} & 0 & -i\sqrt{2}t_{y} & 0 & 0 & 0\\
0 & \frac{(i-1)\gamma}{\sqrt{2L_{x}L_{z}}} & 0 & -i\sqrt{2}t_{y} & 0 & 0\\
0 & \frac{(i-1)\gamma}{\sqrt{2L_{x}L_{z}}} & 0 & 0 & -i\sqrt{2}t_{y} & 0\\
\frac{(1+i)\gamma}{\sqrt{2L_{x}L_{z}}} & 0 & 0 & 0 & 0 & -i\sqrt{2}t_{y}
\end{array}\right)
\end{multline*}
Finally, Eq. \eqref{eq:FisherLee}
yields
\begin{equation}
S=\left(\begin{array}{cccc}
\frac{1}{i+\Phi} & 0 & \frac{\Phi}{i+\Phi} & 0\\
0 & \frac{1}{i-\Phi} & 0 & \frac{\Phi}{-i+\Phi}\\
\frac{\Phi}{i+\Phi} & 0 & -\frac{1}{i+\Phi} & 0\\
0 & \frac{\Phi}{-i+\Phi} & 0 & \frac{1}{-i+\Phi}
\end{array}\right)\label{eq:SL(x)}
\end{equation}
where $\Phi=\frac{t_y AL_{z}\phi_{x}}{\sqrt{2}|\gamma|^{2}}$. The resulting Eigenvalue
phase winding is compared to the full-scale numerical calculation in Fig. \ref{fig:phase_winding_plot}, the excellent agreement between both curves quantitatively confirms the model leading to Eq. \eqref{eq:SL(x)}. For larger coupling strength $\gamma$ [\textit{i.e.} $\gamma=5$ as used in the numerics for Figs. \ref{fig:lead and flux}(b) and \ref{fig:phase diagram}], the assumption $H_{\mathrm{S}} \simeq \bar{H}_{y}^\mathrm{STI}$ becomes invalid as surface states strongly hybridize with the lead and can no longer be labeled with surface momenta. Accordingly, Eq. \eqref{eq:SL(x)} then deviates from the full numerical solution.

The phase winding shown in Fig. \ref{fig:phase_winding_plot} (STI, $\phi_x:0\rightarrow \pi$ and $\phi_z=0$) is nontrivial. In a similar fashion, all other phase windings in the absence of disorder can be modeled using the effective low-energy and agree with the \noun{kwant} results. In general, a surface momentum trajectory that leaves or enters an odd number of surface Dirac points corresponds to a non-trivial phase winding.  In the following, as we discuss the only case where two Dirac points are reached for the same flux configuration, we show why we
prefer using the extended lead $H_{\mathrm{L}}$ {[}Eq. \eqref{eq:HL}{]} instead
of the strictly one-dimensional lead $H_{\mathrm{L}}^{\prime}$ [Eq. \eqref{eq:thin lead}]. 
\begin{figure}
\centering{}\includegraphics[scale=1]{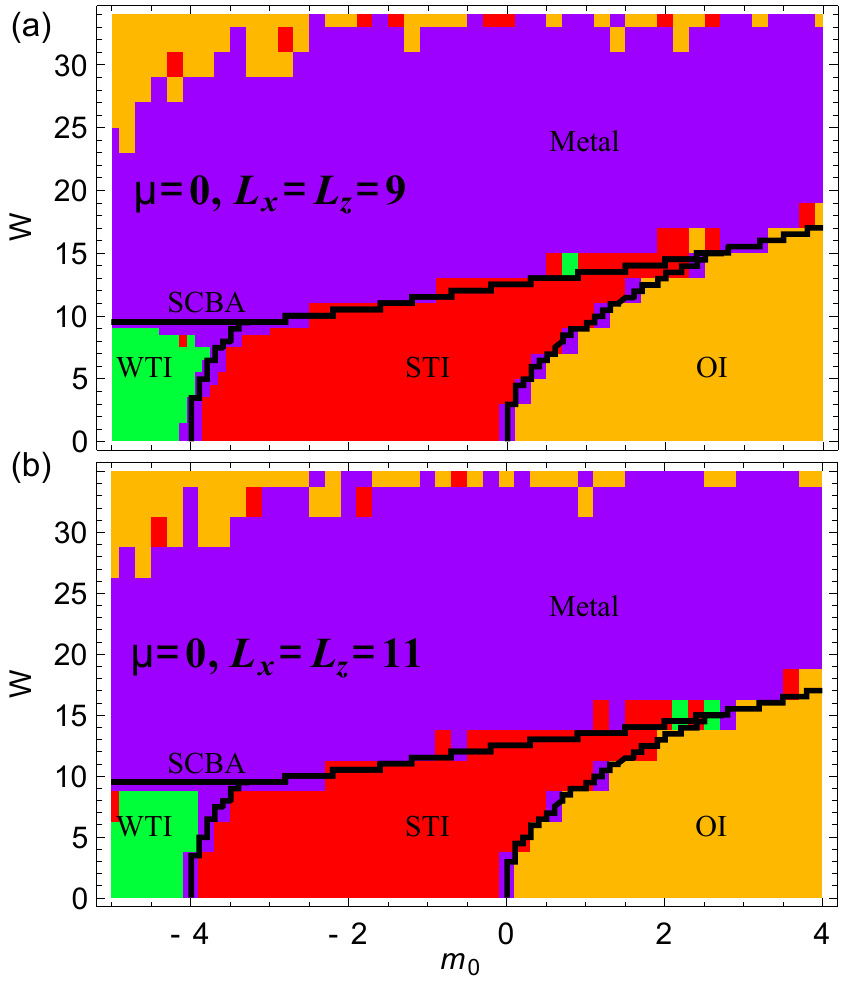}
\caption{\label{fig:FiniteSize}Comparison of topological phase diagrams for
(a) $L_{x,z}=9$ and (b) $L_{x,z}=11$ which show excellent agreement.
For the larger system, the resolution in parameter space is reduced.
SCBA phase boundaries are included to facilitate comparison. }
\end{figure}

\subsection{Motivation for an extended lead}

Consider the situation $m_0<-4$ and even system dimensions. For $\phi_z=0$ and $\phi_x:0\rightarrow \pi$ the trajectories of surface momenta simultaneously leave the two Dirac cones at $\bar{\mathbf{Q}}_{1,2}=(0,\pi)$ and $(\pi,0)$, respectively. The effective surface Hamiltonian is
\begin{equation}
\bar{H}_{y}^\mathrm{WTI}\left(\phi_{x}\right)=2\left(\begin{array}{cc}
\begin{array}{cc}
-\phi_{x} & 0\\
0 & \phi_{x}
\end{array} & 0\\
0 & \begin{array}{cc}
\phi_{x} & 0\\
0 & -\phi_{x}
\end{array}
\end{array}\right)\label{eq:H_L2}
\end{equation}
with basis states in Eqs. \eqref{eq:STI_surface_state_spin_up_around-0-1} and \eqref{eq:STI_surface_state_spin_down_around-0-1} for $\bar{\vq}_j\simeq \bar{\vQ}_j$, $j=1,2$. Now consider a lead which is weakly coupled to just a single site at the surface of the system, say at $\bar{\vr}=(0,0)$, and calculate the Green function $G^R=(-\bar{H}_{y}^\mathrm{WTI}-W^\prime_\mathrm{L}-\Sigma_{\mathrm{L}}^\prime)^{-1}$. Crucially, the coupling matrix elements (denoted by $\Gamma^\prime$ in the following) for the two different surface Dirac cones $j=1,2$ are identical in such a situation since they fail to resolve the different in-plane momenta of the surface states. Representing the 2x2 blocks of Eq. \eqref{eq:H_L2} by $\pm h$ we obtain generically
\begin{equation}
G^{R}=\left(\begin{array}{ccc}
-h & 0 & -\Gamma^\prime\\
0 & h & -\Gamma^\prime\\
-\Gamma^{\prime \dagger} & -\Gamma^{\prime \dagger} & -\Sigma_{L}^\prime
\end{array}\right)^{-1}
\end{equation}
where (after matrix inversion) the relevant on-site part at $y=-1$ is just  $-1/\Sigma_{L}$, leading to a scattering matrix independent of $\phi_x$. This trivial phase winding is consistent with the discussion in Sec. \ref{sec:3}. However, any small perturbation that acts differently on the two Dirac cones invalidates the exact cancellations and causes a steep but still trivial phase winding that is increasingly harder to track for a decreasing perturbation strength. In numerical practice, finite precision of the arithmetics plays the role of a tiny perturbation which prevents proper eigenvalue phase tracking. Although even a small amount of disorder ($W=0.1$) is a sufficiently strong perturbation to overcome the problem, an improved lead and lead-system coupling than can distinguish between the two surface Dirac cone basis states are desirable.

A lead which is extended in, say,
$z$ direction {[}see Fig. \ref{fig:RoleOfLead}(b){]} can carry modes
that probe the different in-plane momenta of surface states. Such a lead is realized by our
default choice $H_{\mathrm{L}}$ in Eq. \eqref{eq:HL}. The modes are proportional to $e^{i0z}$ or $e^{i\pi z}$ and are thus mutually orthogonal to the the surface modes if these 
belong to Dirac cones with $Q_z=0$ or $\pi$.  Thus, the
scattering scenario described in this section becomes an effective double copy of the scenario
in Subsection \ref{subsec:STI}. Now, the steepness of the (double) phase winding is conveniently
controlled by $\gamma$, which justifies the increased numerical cost due
to the doubling of scattering channels. 

\section{Finite-size effects}
\label{app:b}

Figure \ref{fig:FiniteSize} proves the successful
suppression of finite-size effects for the system dimensions reached
in this work. The phase diagram of Sec. \ref{sec:PhaseDiagram} remains
unchanged if we increase $L_{x,z}$ from $9$ to $11$ (and thus the
volume by $50\%$).


\end{document}